# MACHINE LEARNING IN NETWORK SECURITY USING KNIME ANALYTICS

MuntherAbualkibash

School of Information Security and Applied Computing, College of Technology, Eastern Michigan University, Ypsilanti, MI, USA

## ABSTRACT

*Machine learning has more and more effect on our every day's life. This field keeps growing and expanding into new areas. Machine learning is based on the implementation of artificial intelligence that gives systems the capability to automatically learn and enhance from experiments without being explicitly programmed. Machine Learning algorithms apply mathematical equations to analyze datasets and predict values based on the dataset. In the field of cybersecurity, machine learning algorithms can be utilized to train and analyze the Intrusion Detection Systems (IDSs) on security-related datasets. In this paper, we tested different machine learning algorithms to analyze NSL-KDD dataset using KNIME analytics.*

## 1. INTRODUCTION

In today's connected world, where billions of people access the internet, anything that depends on the internet for communication, or is connected to a computer or any type of smart device, can be affected by several kinds of cyber attacks. As a result, many organizations, either public or private, have to deal with continuous and complicated different types of cyber attacks and cyber threats. The fact that cyber threats and cyber attacks now permeate every facet of society shows why cybersecurity is crucially important.

Cybersecurity is the action of securing programs, networks, and systems from any cyberattacks. These cyberattacks are mainly intended at gaining access to, altering, or deleting critical data; stealing money from users; or stopping usual business operations.

Intrusion Detection System (IDS) is one of the important and dynamic areas to handle cyberattacks.IDS is an implementation or a technique which can detect an attack attempt by analyzing the activity of network or system then IDS will raise the alarm. Any system that can decide for further steps is named Intrusion Prevention System (IPS). The role of IDS is to raise the security level by identifying malicious and suspicious events that could be detected in a computer or network system.

One of the most popular study areas in intrusion detection is anomaly-based and signature-based detection. Anomaly-based intrusion detection talks about the case of detecting untypical events in the network traffic that do not follow the normal patterns. It is presumed that anything that is untypical or, in another word, anomalous could be critical and to some extent associated with some security events. Signature-based intrusion detection tasks is to detect attacks by looking for specific patterns where these detected patterns are referred to as signatures.[1] This paper is organized as follows: Section two gives a brief introduction about Intrusion Detection Systems (IDSs). Section three talks about NSL-KDD dataset. Section four summarizes the tested machine learning algorithms. Section five presents the experiments and results. The last section is the conclusion.

## 2. INTRUSION DETECTION SYSTEMS(IDSS)

### 2.1. Intrusion Detection Systems Categories

IDSs can be recognized based on two different categories:

- Host-based Intrusion Detection Systems (HIDS): They can scan and watch the system actions on which it has been installed. HIDS might detect any changes in the integrity of files of any file system and analyze log files to look for any malicious or suspicious activity.

- Network-based Intrusion Detection Systems: They focus on scanning and watching the network infrastructure. They analyze the packets flow of the network and examining packets' headers and contents to detect any possible attack on the network.

- As mentioned before, two different actions are applied by both types of IDS to analyze data:
  Signature-based Intrusion Detection Systems: Such systems can detect known attacks by comparing them with stored patterns, however, it can't recognize new attacks. So the detection will be based on signatures of known attacks as well as any rules determined by a system administrator.
- Anomaly-based Intrusion Detection Systems: Such systems can create a model based on normal system activity, and then use the built model to evaluate observed activity to determine if the observed activity is an anomaly.

### 2.2. Anomaly Detection Methods

There are many algorithms and methods from different classes used by researchers to implement anomaly detection in network traffic. Some techniques are implemented based on a statistical point of view where statistics are used to compute and to determine if the observed case is an anomaly[2]. Other techniques are based on machine learning algorithms that are applied to detect an anomaly.

Machine learning methods can be classified into three categories:

- Supervised learning: A training set have labelled examples and an algorithm will match a new observation with just one class.
- Unsupervised learning: Training set does not have labels or any details about a possible group in it. In the time of training, the algorithm sets groups and determines its level of similarity.
- Semi-supervised learning: Training set has a small amount of labelled example with a large number of unlabelled examples. In other words, semi-supervised learning takes place between supervised learning, where training data is labelled, and unsupervised learning, where training data is not labelled.

Algorithms from the first two categories have been implemented in a network anomaly detection problem. It is expected that attacks can be detected since they are abnormal events and will be classified by the algorithm model. Models are used by machine learning algorithms to analyze network traffic.

## 3. DATASETS

### 3.1. NLS-KDD Dataset

Over the last decades, a few datasets have been utilized to examine network anomaly detection systems. The most well-known dataset is KDDcup99. This dataset contains about 4,900,000 samples where 300,000 represent 24 different attack types. Every sample is described by 41 features and labelled as either an attack or normal. However, KDDcup99 has been criticized by many researchers because it has many redundant records and irregularities[3, 4].

To fix this problem, a new dataset, NSL-KDD was proposed, that contains selected records of the entireKDDcup99 dataset. The differences between KDDcup99 and NSL-KDD are that the redundant records in NSL-KDD have been deleted from the training set to make sure that the built classifiers are not biased. Also, duplicated records have been excluded to have a better detection rate once applied to some methods. As a result, it is highly recommended to stop using the KDDcup99 dataset and to use the NSL-KDD dataset instead to evaluate machine learning algorithms because it solves the issues in the KDDcup99.

#### 3.1.1. Attacks categories in the NSL-KDD dataset

There are four attacks categories represented in the NSL-KDD dataset:

1. A denial-of-Service attack (DoS) is an attack intended to freeze or powering down a machine or network by forcing it to be unreachable to its legitimate users. DoS attacks achieve this by flooding the target with traffic or transmitting it information that causes a crash. As a result, the DoS attack blocks intended users of the service or resource they looking for.

2. User to Root Attack (U2R) is an attack where the attacker logs in on the system with a normal user account then make the effort to find any vulnerability in the system and use it to obtain the root or the admin privileges.

3. Remote to Local Attack(R2L)is an attack where the attacker sends packets to a targeted machine in which the attacker does not have access to it to expose any vulnerability in the targeted the machines and exploit privileges that only a local user can have on that machine.

4. Probing Attack is an attack where the attacker scans a targeted machine as a means to find any vulnerability on that machine that can be used to compromise the system.

## 4. MACHINE LEARNING ALGORITHMS

There are many techniques in IDS are built on machine learning approaches. This paper will go over several machine learning algorithms that have been used in the cybersecurity research area.

## 4.1. Decision Trees

Decision Trees algorithms are one of the used algorithms to solve classification where algorithms sort data into classes, like whether an event is an attack or not. Decision Trees are made up of nodes, branches, and leaves where every node presents as an attribute or feature and every branch present as a rule or decision, and every leaf presents as an outcome. We can look at the decision tree as a series of yes/no questions applied to our data resulting in a predicted class.

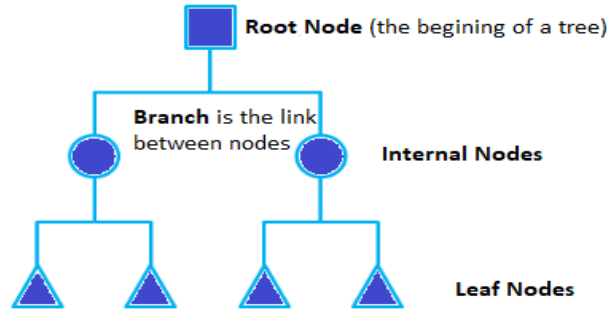

Figure 1. Decision Trees of two levels.

There are many algorithms derived from the decision tree algorithm. Some of them are ID3, C4.5, J48, etc. The issue with ID3 algorithm is that the information might be overfitted.C4.5 is an enhanced version of ID3 and it handles the overfitting issue. J48is an open-source of C4.5.

## 4.2. Random Forest

The random forest is a model structured from many decision trees. This model applies two key concepts that yield it the name random:

- Random sampling of training data will be considered when creating trees, where each tree in a random forest learns from a random sample of the training data. The purpose behind training each tree on different samples, even though each tree may have high variance regarding a specific set of the training data, is that the entire forest will have lower variance without increasing the bias. During testing, predictions are formed by taking the predictions averages of each decision tree. These steps of training each learner on various bootstrapped subsets of the training data and then taking the predictions averages are called bagging, that is short for bootstrap aggregating.

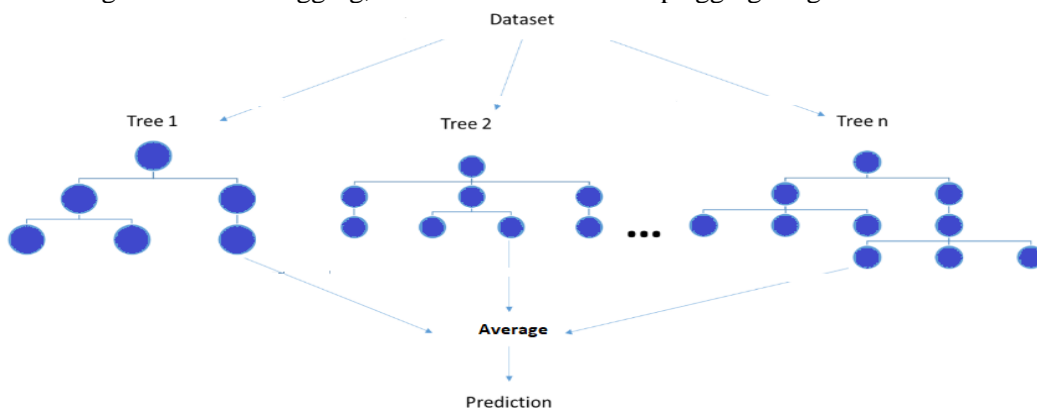

Figure 2. Random forests and decision trees

- Random subsets of features will be considered when dividing nodes in each decision tree. For example, if there are 4 features, at each node in each tree, only 2 random features will be taken into consideration for dividing the node.

### 4.3. Naive Bayes

Naive Bayes is a classifier where the machine learning model is built based on probability to be used in classification using the Bayes theorem. For example, by applying Bayes theorem, we can calculate the probability of an attack to happen when an event has occurred.

### 4.4. Support Vector Machines (SVM)

Support vector machines are a supervised learning algorithm that can be applied in classification cases. It is usually applied to a small dataset because it takes a long time to process. SVM is built based on the concept of determining a hyperplane that best divides the features into several domains. For example, if we want to create a function (hyperplane) that will identify the two cases, such that whenever a server received so many packets is it going to be classified as an attack or normal? The following are the figures of two scenarios in which the hyperplane are drawn.

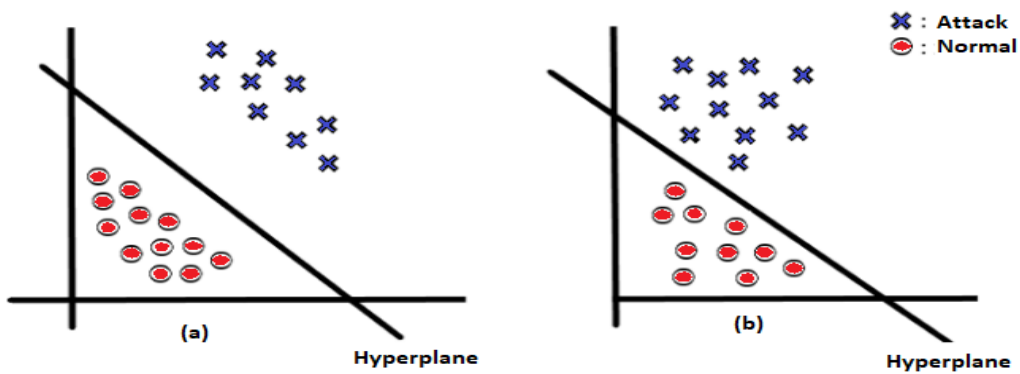

Figure 3. SVM will come up with an optimal hyperplane that will classify the different classes

The points near the hyperplane are called: support vector points and the space between the vectors and the hyperplane are called: margins.

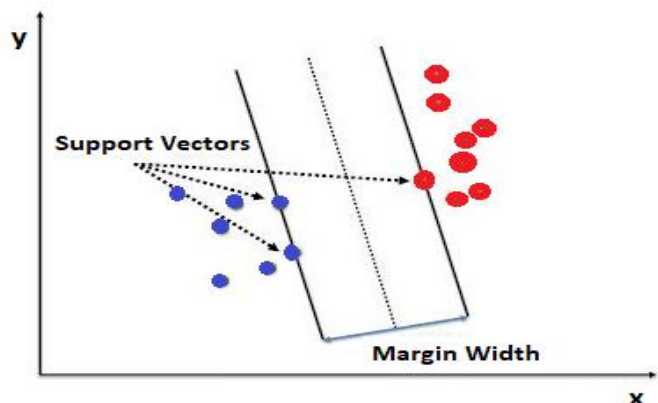

Figure 4. Support vector points andmargins

## 4.5. Artificial Neural Network

Artificial neural networks are one of the widely used algorithms in machine learning. They are brain-inspired systems that are designed to copy the way how humans learn. Neural networks contain input and output layers and also, it can have hidden layers that convert the input into something that the output layer can utilize. In Artificial Neural Network, if more data is used as a training set then it will become more accurate. It is similar when someone keeps doing a task over and over. Over time, that person becomes more efficient as well as makes a small number of mistakes.

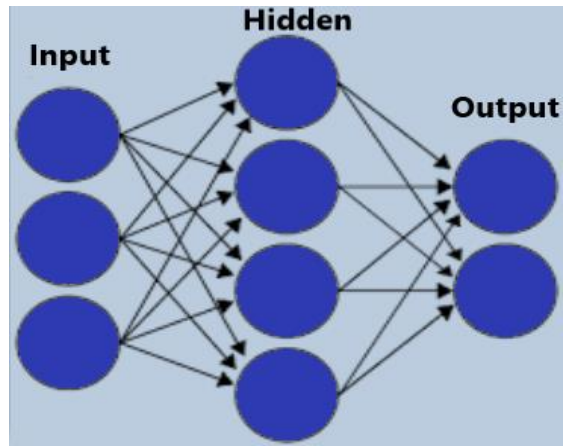

Figure 5. A simple neural network with a simple hidden layer

## 4.6. Gradient Boosted Trees

The Gradient boosted trees used an ensemble of several trees to make more powerful prediction models for classification by building a series of trees where each tree is trained to try correcting mistakes in the previous tree in the series.

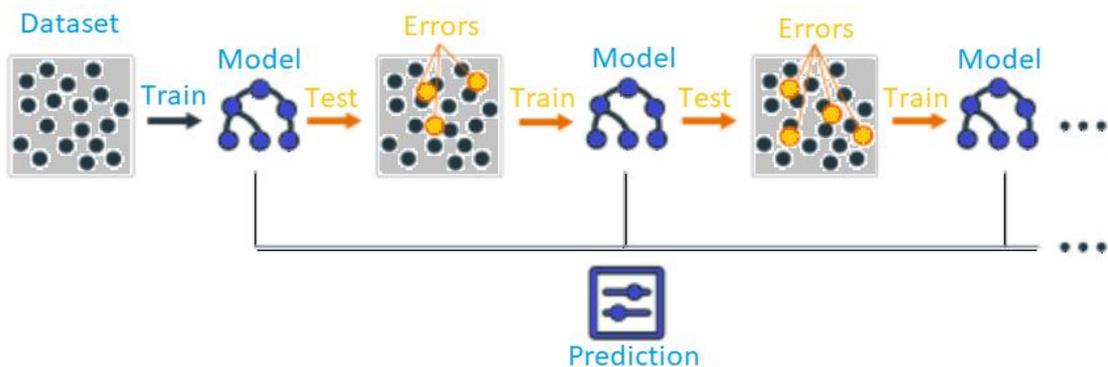

Figure 6. Add new models to the ensemble sequentially

## 4.7. k-Nearest Neighbor (kNN)

The k-Nearest-Neighbors algorithm of classification is one of the straight forward algorithms in machine learning. The classification relies on identifying similar data points in the training data then making a prediction based on their classifications. kNN is one of the algorithms that give better results on a small size data-sets that do not have several features.

## 5. EXPERIMENTS AND RESULTS

Using machine learning algorithms in Intrusion Detection System (IDS) is an attracting research area for cyber security researchers around the world [5-14]. In this research, experiments were made based on classifying NSL-KDD dataset to either normal or attack. NSL-KDD dataset was downloaded from the webpage of the Canadian Institute for Cybersecurity (CIC) that is based at the University of New Brunswick [15]. Subset files from the NSL-KDD dataset were used: KDDTrain+.arff and KDDTest+.arff.

- KDDTrain+.arff has a dataset that is used for training purposes and the data is labelled as either normal or anomaly. This file is saved in ARFF format.

- KDDTest+.arff has a dataset that is used for testing purposes and the data is labelled as either normal or anomaly. This file is saved in ARFF format.

The KDDTest+ dataset contains known and new attacks. New attacks are the ones that do not exist in KDDTrain+ dataset.

### 5.1. Data Pre-processing

Data pre-processing is a crucial step in the journey of making a machine learning model. We have to do some preparation to make sure that we build our machine learning models without any issues. If there is no data pre-processing then our machine learning model won't work properly.

In any dataset, we need to distinguish something very important which is the difference between the independent variables and the dependent variables. In any machine learning algorithm or any model, independent variables are used to predict a dependent variable e.g., normal or anomaly. Also, we have to deal with cases where we have some missing data in the dataset. The most common idea to handle missing data in a column is to take the mean of that column. In other words, the mean of all the values in a column will replace the missing data in that column.

Furthermore, sometimes in the dataset, there will be a kind of variables that are called categorical variables because simply they contain categories. Since machine learning models are based on mathematical equations, therefore this would cause some problem if we keep the categorical variables, e.g., text, in the equations because we would only want numbers in the equations. As a result, we need to encode the categorical variables into numbers.

The next step features scaling, which is very important in machine learning when variables are not on the same scale, this will cause some issues in some machine learning models.

### 5.2. Classification

Tables 1 and 2 show the results of applying different classification algorithms using KNIME to analyze the NSL-KDD dataset. KNIME is an open-source data analytics. It has a shapely graphical user interface that is easy to use and understand.

We tested several KNIME's machine learning algorithms using two different approaches: In the first approach, the KDDTrain+ has partitioned to 70% training set and 30% testing set so each algorithm is trained on the 70% portion and tested on the 30% portion.

In the second approach, each algorithm is trained using the entire KDDTrain+ and tested using the KDDTest+ dataset. The KDDTest+ dataset contains known and new attacks. New attacks are the ones that do not exist in KDDTrain+ dataset.

The tested machine learning algorithms will generate alerts that can be categorized as follows.

- True positive which means an attack is predicted as an attack.
- False-positive which means a normal network packet is predicted as an attack.
- True negative which means a normal network packet is predicted as normal.
- False-negative which means an attack is predicted as normal.

## 5.3. Results

When KDDTrain+ is partitioned to 70% training set and 30% testing set and the 30% portion dataset is used for testing the algorithms, it has been noticed that the majority of the tested algorithms achieved a very high accuracy as shown in Table 1.

However, when KDDTest+ dataset is used, which contains both old and new attacks, a sudden decrease in the accuracy has been noticed on all the tested algorithms. The best accuracy achieved is 79% with Naive Bayes then 78% with Probabilistic Neural Network and Gradient Boost Tree algorithms. The result obviously shows that some of the tested machine learning algorithms are useful for detecting attacks on which they are trained but performs poorly on new attacks as shown in Table 2.

Table 1. The accuracy statistics on KDDTrain+ Dataset that is partitioned to 70% training set and 30% testing set.

| Machine Learning Algorithm | Class | True Positives | False Positives | True Negatives | False Negatives | Accuracy |
|---|---|---|---|---|---|---|
| Decision Tree | Attack | 17537 | 30 | 20191 | 34 | 0.998 |
| | Normal | 20191 | 34 | 17537 | 30 | |
| Random Forest | Attack | 17532 | 20 | 20201 | 39 | 0.998 |
| | Normal | 20201 | 39 | 17532 | 20 | |
| Naive Bayes | Attack | 18324 | 2635 | 14936 | 1897 | 0.880 |
| | Normal | 14936 | 1897 | 18324 | 2635 | |
| Support Vector Machine (SVM) | Attack | 16527 | 655 | 19566 | 1044 | 0.955 |
| | Normal | 19566 | 1044 | 16527 | 655 | |
| Probabilistic Neural Network (PNN) | Attack | 18467 | 807 | 16764 | 1754 | 0.932 |
| | Normal | 16764 | 1754 | 18467 | 807 | |
| Gradient Boosted Trees | Attack | 20183 | 57 | 17514 | 38 | 0.997 |
| | Normal | 17514 | 38 | 20183 | 57 | |
| K Nearest Neighbor | Attack | 20145 | 78 | 17493 | 76 | 0.996 |
| | Normal | 17493 | 76 | 20145 | 78 | |

Table 2.  The accuracy statistics on KDDTest+ Dataset.

| Machine Learning Algorithm | Class | True Positives | False Positives | True Negatives | False Negatives | Accuracy |
|---|---|---|---|---|---|---|
| **Decision Tree** | Attack | 7270 | 212 | 9499 | 5563 | 0.74 |
| | Normal | 9499 | 5563 | 7270 | 212 | |
| **Random Forest** | Attack | 7106 | 645 | 9066 | 5727 | 0.72 |
| | Normal | 9066 | 5727 | 7106 | 645 | |
| **Naive Bayes** | Attack | 8652 | 476 | 9235 | 4181 | 0.79 |
| | Normal | 9235 | 4181 | 8652 | 476 | |
| **Support Vector Machine (SVM)** | Attack | 7332 | 779 | 8932 | 5501 | 0.72 |
| | Normal | 8932 | 5501 | 7332 | 779 | |
| **Probabilistic Neural Network (PNN)** | Attack | 8630 | 712 | 8999 | 4203 | 0.78 |
| | Normal | 8999 | 4203 | 8630 | 712 | |
| **Gradient Boosted Trees** | Attack | 8145 | 218 | 9493 | 4688 | 0.78 |
| | Normal | 9493 | 4688 | 8145 | 218 | |
| **K Nearest Neighbor** | Attack | 7172 | 693 | 9018 | 5661 | 0.72 |
| | Normal | 9018 | 5661 | 7172 | 693 | |

The following figures show how to use KNIME workflow to build and run different KNIME's machine learning algorithms:

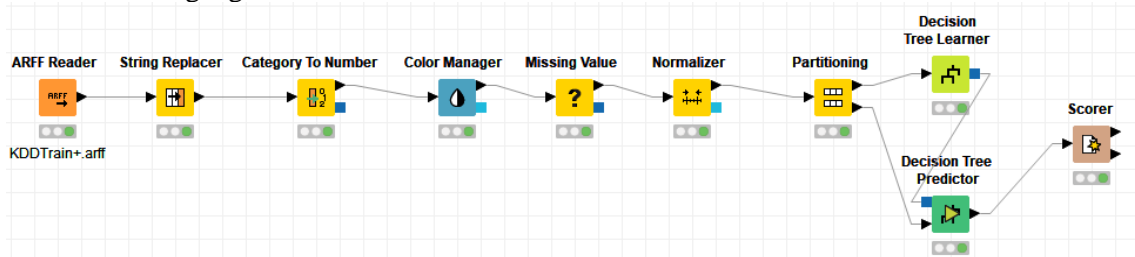

Figure 7. KNIME workflow using Decision Tree Learner where KDDTrain+ Dataset is partitioned to 70% training set and 30% testing set

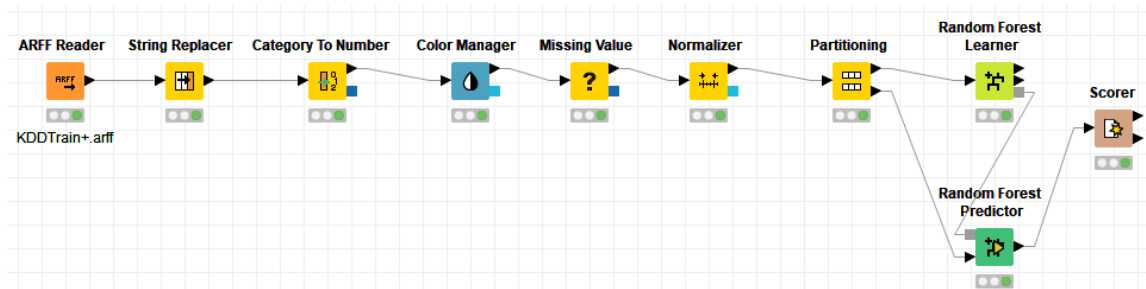

Figure 8. KNIME workflow using Random Forest Learner where KDDTrain+ Dataset is partitioned to 70% training set and 30% testing set

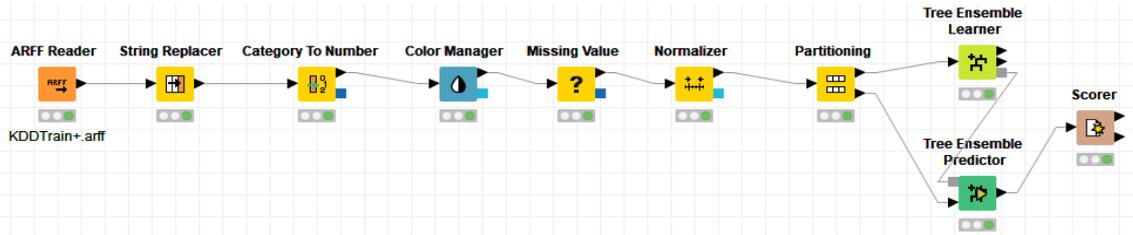

Figure 9. KNIME workflow using Tree Ensemble Learner where KDDTrain+ Dataset is partitioned to 70% training set and 30% testing set

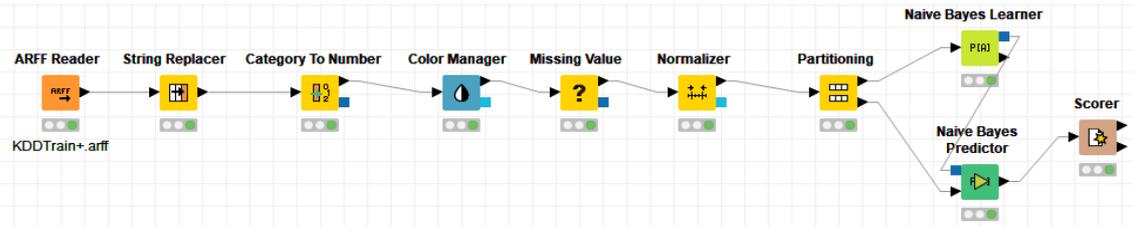

Figure 10. KNIME workflow using Naive Bayes Learner where KDDTrain+ Dataset is partitioned to 70% training set and 30% testing set

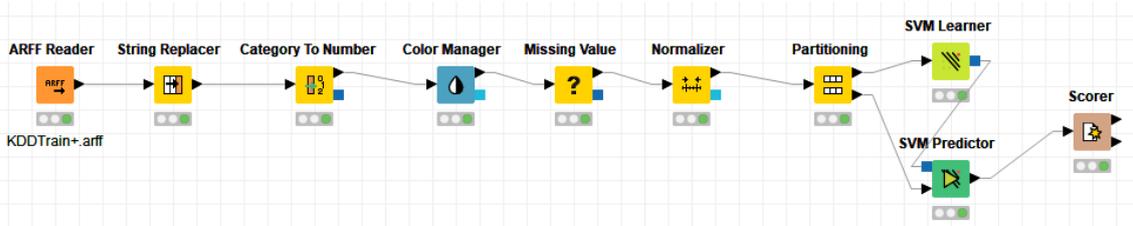

Figure 11. KNIME workflow using SVM Learner where KDDTrain+ Dataset is partitioned to 70% training set and 30% testing set

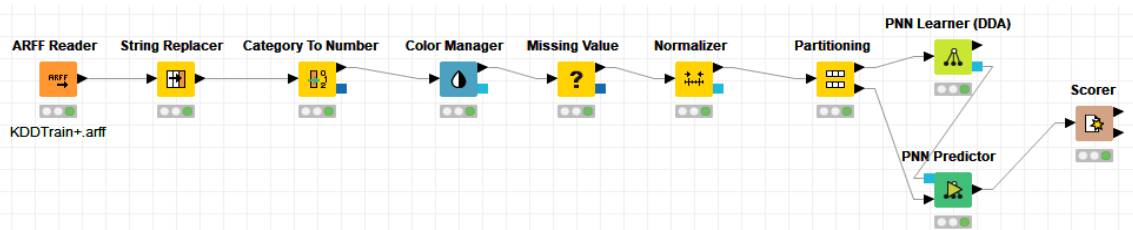

Figure 12. KNIME workflow using PNN Learner where KDDTrain+ Dataset is partitioned to 70% training set and 30% testing set

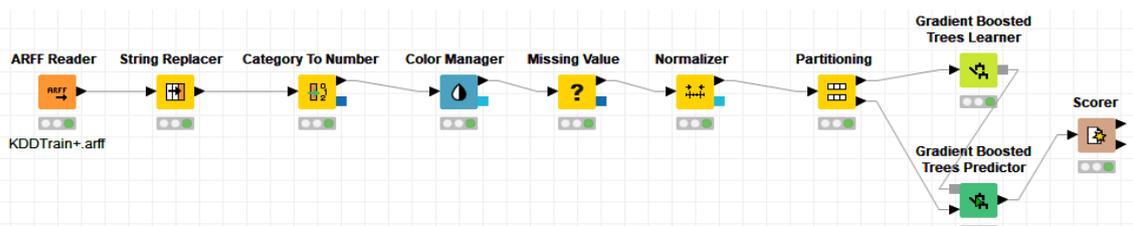

Figure 13. KNIME workflow using Gradient Boosted Trees Learner where KDDTrain+ Dataset is partitioned to 70% training set and 30% testing set

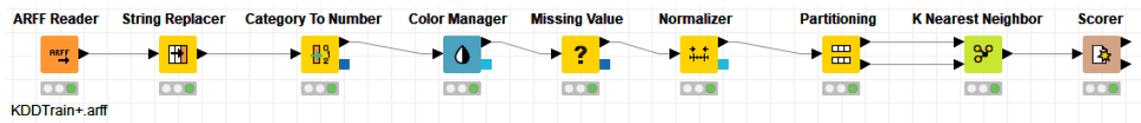

Figure 14. KNIME workflow using K Nearest Neighbor where KDDTrain+ Dataset is partitioned to 70% training set and 30% testing set

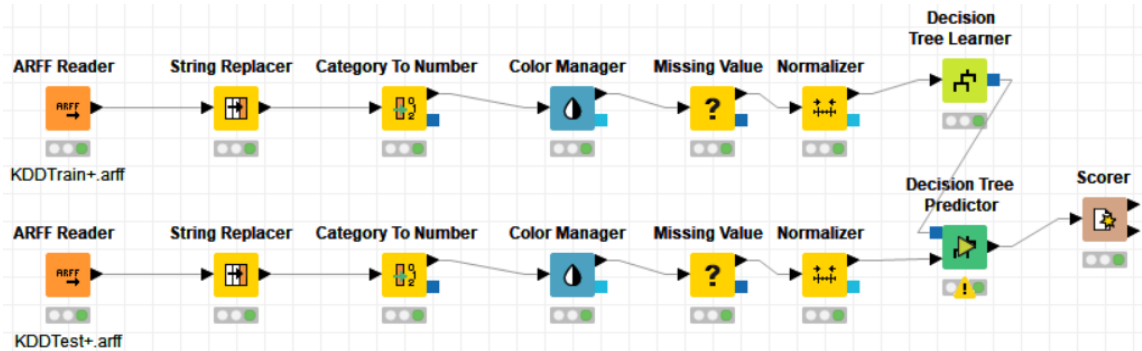

Figure 15. KNIME workflow using Decision Tree Learner on KDDTest+ Dataset

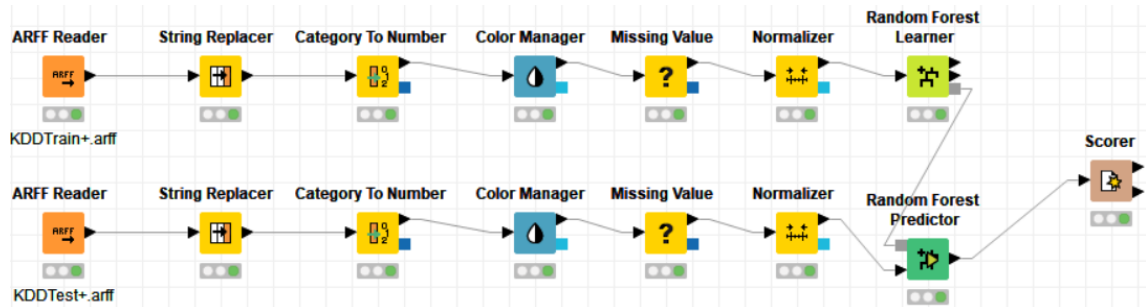

Figure 16. KNIME workflow using Random Forest Learner on KDDTest+ Dataset

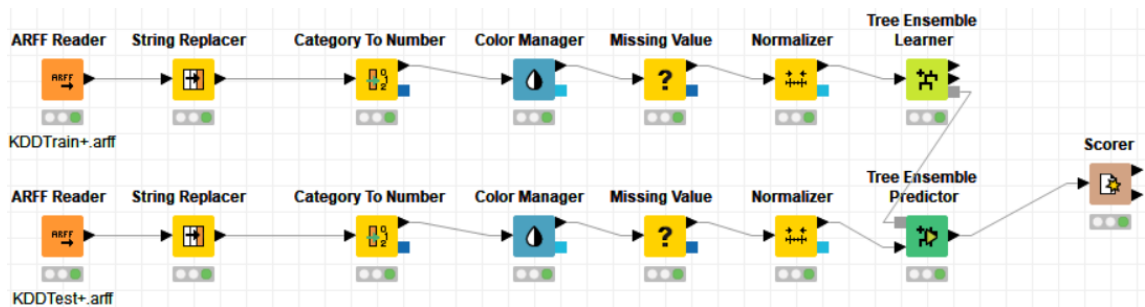

Figure 17. KNIME workflow using Tree Ensemble Learner on KDDTest+ Dataset

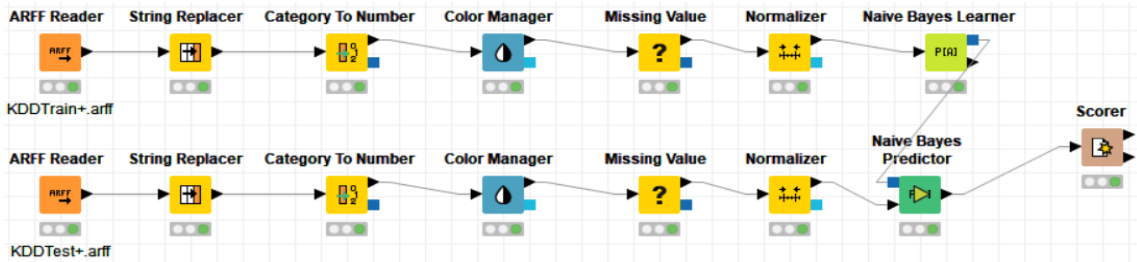

Figure 18. KNIME workflow using Naive Bayes Learner on KDDTest+ Dataset

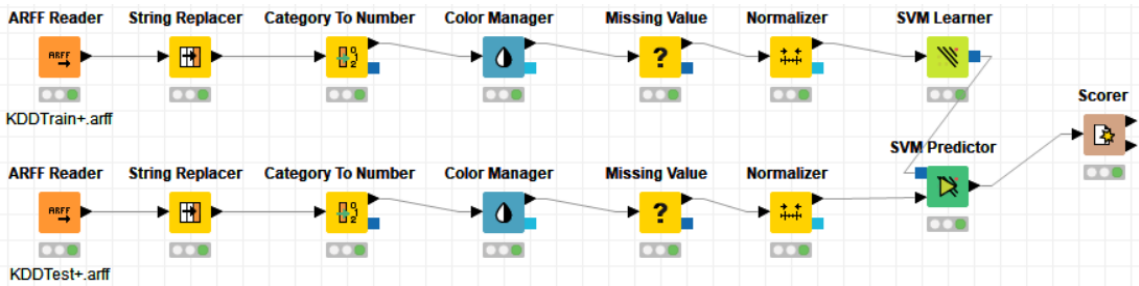

Figure 19. KNIME workflow using SVM Learner on KDDTest+ Dataset

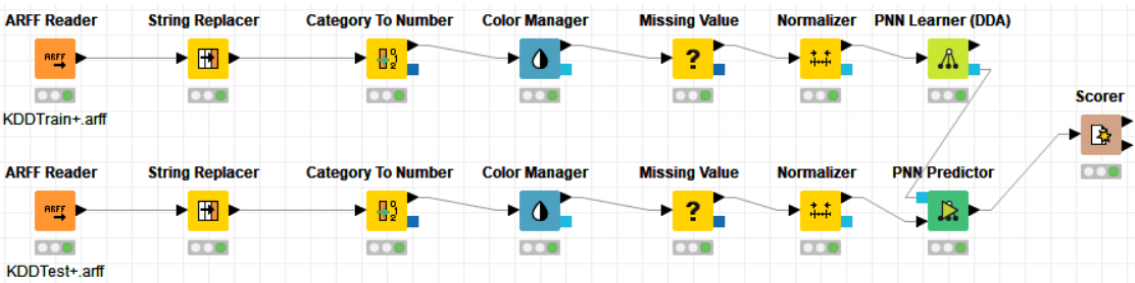

Figure 20. KNIME workflow using PNN Learner on KDDTest+ Dataset

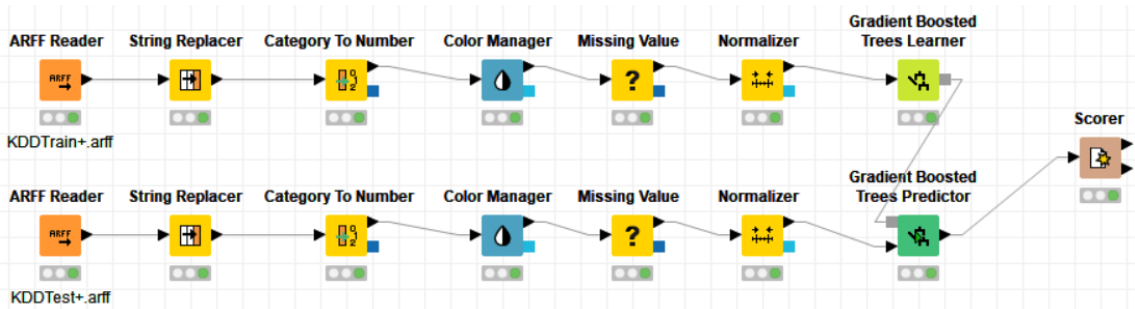

Figure 21. KNIME workflow using Gradient Boosted Trees Learner on KDDTest+ Dataset

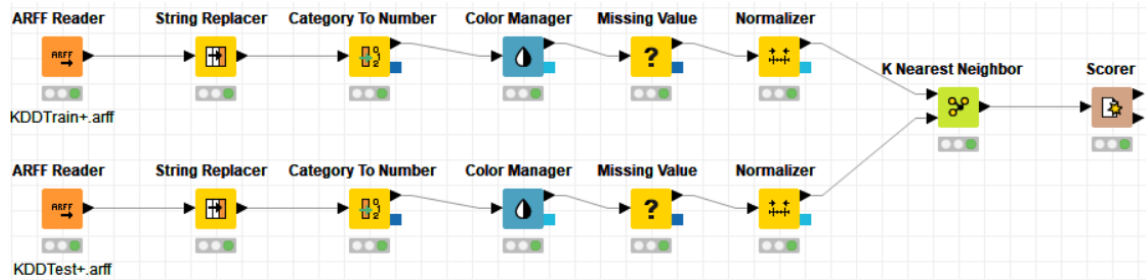

Figure 22. KNIME workflow using K Nearest Neighbor on KDDTest+ Dataset

# 6. CONCLUSIONS

In this paper, different KNIME's machine learning algorithms that can be used in intrusion detection systems (IDSs) have been tested to analyze the NSL-KDD dataset. Also, an accuracy comparison of these algorithms is given. Every algorithm has its features that have a significant role in enhancing IDSs when compared to other algorithms. The results indicate that almost most of the tested machine learning algorithms used in this paper show excellent performance on detecting attacks using a dataset on which they are trained. However, the performance of the algorithms goes down when the testing dataset includes new attacks.